\documentclass[12pt]{article}

\usepackage{graphicx}
\begin{document}

\begin{center}
{\bf The R\'{e}nyi entropy and entropic cosmology} \\
\vspace{5mm} S. I. Kruglov
\footnote{E-mail: kruglov@rogers.com}
\underline{}
\vspace{3mm}

\textit{Department of Physics, University of Toronto, \\60 St. Georges St.,
Toronto, ON M5S 1A7, Canada\\
Canadian Quantum Research Center, \\
204-3002 32 Ave., Vernon, BC V1T 2L7, Canada} \\
\end{center}

\begin{abstract}
Entropic cosmology with the R\'{e}nyi entropy of the apparent horizon $S_R=(1/\alpha)\ln(1+\alpha S_{BH})$, where $S_{BH}$ is the Bekenstein--Hawking entropy, is studied. By virtue of the thermodynamics-gravity correspondence a model of dark energy is investigated. The generalised Friedmann equations for the Friedmann--Lema\^{i}tre--Robertson--Walker spatially flat universe with the barotropic matter fluid are obtained. We compute the dark energy density $\rho_D$, pressure $p_D$ and the deceleration parameter $q$ of the universe. At some model parameters the normalized density parameter of the matter $\Omega_{m0}\approx 0.315$ and the deceleration parameter $q_0\approx -0.535$ for the current epoch, which are in the agreement with the Planck data, are found. Making use of the thermodynamics-gravity correspondence, we describe the  late time of the universe acceleration. The entropic cosmology considered is equivalent to cosmology based on the teleparallel gravity with the definite function $F(T)$. The Hubble parameters are in approximate agreement (within $5$ percents) with the observational Hubble data for redshifts $0.07\leq z \leq 1.75$ at the entropy parameter $\alpha\approx 0.305~GH_0^2$.
\end{abstract}

\section{Introduction}

Data from observations of type Ia supernovae (SNe Ia) \cite{Riess0, Perlmutter} and the
cosmic microwave background (CMB) anisotropies \cite{Aghanim},  and baryon acoustic oscillations (BAO) \cite{Adame} shown that the universe currently accelerates. Thus, observations provide evidence for a dark-energy component. The current universe acceleration can be explained by introduction the cosmological constant $\Lambda$ in the Einstein--Hilbert action. Then there will be the contribution to the energy density in the Friedmann equation. Such energy density, named the dark energy, leads to the current universe acceleration and large scale homogeneity and isotropy. But according to the observational data there is the deviation from such scenario with the constant cosmological constant in the favor of dynamical cosmological constant. Thus, there are discrepancies in the determination of the Hubble constant from different sources (Cepheids, supernovae Ia, and CMB) \cite{Lusso,Valentino,Riess,Riess1}.

There is another way to describe the universe acceleration by exploring entropic cosmology based on the thermodynamics of the apparent horizon in a space-time with the Friedmann--Lema\^{i}tre--Robertson--Walker (FLRW) metric \cite{Akbar,Cai,Cai0,Paranjape,Sheykhi,Cai1,Wang,Jamil,Gim,Fan,Agostino,Sanchez} because of a correspondence between gravity and thermodynamics. The entropy of black holes as well as the entropy of the apparent horizon in cosmology is a function of the horizon area, and the temperature is linked with the surface gravity \cite{Bekenstein,Hawking,Jacobson,Padmanabhan,Padmanabhan1,Hayward}. The first law of apparent horizon thermodynamics allows us to obtain Friedmann equations because the apparent horizon for the FLRW spatially flat space-time presents a thermodynamic system \cite{Hayward1,Bak,Akbar,Cai0}. Due to the long-range nature of gravity, nonadditive entropies \cite{Tsallis,Barrow,Renyi,Kaniadakis,Masi,Czinner,Kruglov,Kruglov1} were studied and holographic dark energy models were considered  \cite{Jahromi,Ren,Mejrhit,Majhi,Pavon,Landim,Tamri}.

Here, we explore the R\'{e}nyi entropy to study the entropic cosmology based on the modified Friedmann equations.
The equation of state (EoS) for barotropic perfect fluid with $p=w\rho$, where $p$ is the matter pressure and $\rho$ is the matter energy density, was utilized. We compute the dark energy density and pressure and show that the late time universe acceleration occurs.
For some model parameters, the normalized density parameter of the matter $\Omega_{m}\approx 0.315$ and the deceleration parameter $q_0\approx -0.535$ for the current epoch are obtained, that are in agreement with the Planck data. It is shown that the entropic cosmology considered is equivalent to cosmology within the $F(T)$ teleparallel gravity with a torsion. The values of predicted Hubble parameters are in approximate agreement with the observational Hubble data for redshifts $0.07\leq z \leq 1.75$ within 5 percents.

We utilize units with $\hbar=c=k_B=1$.

\section{The apparent horizon thermodynamics and Friedmann equations}

Let us study the thermodynamics of the apparent horizon in a space-time with the FLRW spatially flat metric which is given by
\begin{equation}
ds^2=-dt^2+a(t)^2(dx^2+dy^2+dz^2),
\label{1}
\end{equation}
were $a(t)$ is a scale factor. In a space-time with the FLRW spatially flat metric the apparent horizon radius coincides with the Hubble radius which is the  distance from an observer where the expansion of the universe causes objects to recede at the speed of light. The apparent horizon radius is defined for $c=1$ as
\begin{equation}
R_h=\frac{1}{H},
\label{2}
\end{equation}
where the Hubble parameter of the universe is $H=\dot{a}(t)/a(t)$ and $\dot{a}(t)=\partial a/\partial t$. To obtain the Friedmann equation within entropic cosmology we consider the first law of apparent horizon thermodynamics which is given by \cite{Hayward,Hayward1,Bak}
\begin{equation}
dE=-T_hdS_h+WdV_h,
\label{3}
\end{equation}
where $W=\frac{1}{2}(\rho-p)$ is the work density, $E=\rho V_h=(4\pi/3)\rho R_h^3$ and $\rho$, $p$ are the energy density and pressure of a matter, correspondingly. The apparent horizon temperature reads \cite{Cai0}
\begin{equation}
T_h=\frac{H}{2\pi}\left|1+\frac{\dot{H}}{2H^2}\right|.
\label{4}
\end{equation}
It was mentioned in \cite{Cai0} that the term with $\dot{H}$ in Eq. (4) is very small. With the help of the continuity equation that represents the conservation law
\begin{equation}
\dot{\rho}=-3H(\rho+p),
\label{5}
\end{equation}
and Eqs. (3), (4), we obtain \cite{Kruglov}
\begin{equation}
\frac{H^3}{2\pi}\dot{S}_h=4\pi (\rho+p).
\label{6}
\end{equation}
According to Eqs. (5) and (6), to obtain the Friedmann equation, we need the entropy function. We utilize here the R\'{e}nyi entropy \cite{Renyi}
\begin{equation}
S_R=\frac{1}{\alpha}\ln(1+\alpha S_{BH}),
\label{7}
\end{equation}
where $S_{BH}=\pi R^2_h/G=\pi/(GH^2)$ is the Bekenstein--Hawking (BH) entropy. The R\'{e}nyi entropy (7) represents the deformation of the BH entropy. To obtain the corrections to BH entropy, we use the series expansion of the entropy (7) for small values of $\alpha S_{BH}$ which is given by
\begin{equation}
S_R=S_{BH}-\frac{\alpha S_{BH}^2}{2}+\frac {\alpha^2S_{BH}^3}{3}+\textit{O}(\alpha^3).
\label{8}
\end{equation}
Equation (8) shows that corrections to the BH entropy $S_{BH}$ decrease the $S_R$ entropy. The similar effect occurs in quantum gravity \cite{Ashtekar}. One can assume that entropy (7) mimics the quantum gravity corrections to $S_{BH}$ entropy. Making use of Eqs. (6) and (7) one obtains the generalized Friedmann equation
\begin{equation}
\frac{\dot{H}H^2}{H^2+b}=-4\pi G(\rho+p),
\label{9}
\end{equation}
where $b=\alpha\pi/G$. The dimension of parameter $b$ is the same as the dimension of $H^2$. In our units ($\hbar=c=k_B=1$) the value of $GH^2$ is dimensionless. Taking into account Eq. (5) and integrating Eq. (9), we find the second generalized Friedmann equation
\begin{equation}
H^2-b\ln\left(\frac{H^2+b}{b}\right)=\frac{8\pi G}{3}\rho.
\label{10}
\end{equation}
If $\alpha=0$ ($b=0$) in Eq. (10), we arrive at the Friedmann equation of General Relativity.
The second term in the left side of Eq. (10) can be treated as a contribution of dark energy to the matter density $\rho$.

\section{The dark energy density, pressure and deceleration parameter}

We represent Eq. (10) in the standard form of Friedmann' equation,
\begin{equation}
H^2=\frac{8\pi G}{3}(\rho+\rho_D),
\label{11}
\end{equation}
where the dark energy density is given by
\begin{equation}
\rho_D=\frac{3b}{8\pi G}\ln\left(\frac{H^2+b}{b}\right).
\label{12}
\end{equation}
We plotted the dimensionless variable $\rho_{D}G/b$ versus $H/\sqrt{b}$ in Fig. 1.
\begin{figure}[h]
\includegraphics [height=2.0in,width=3.5in] {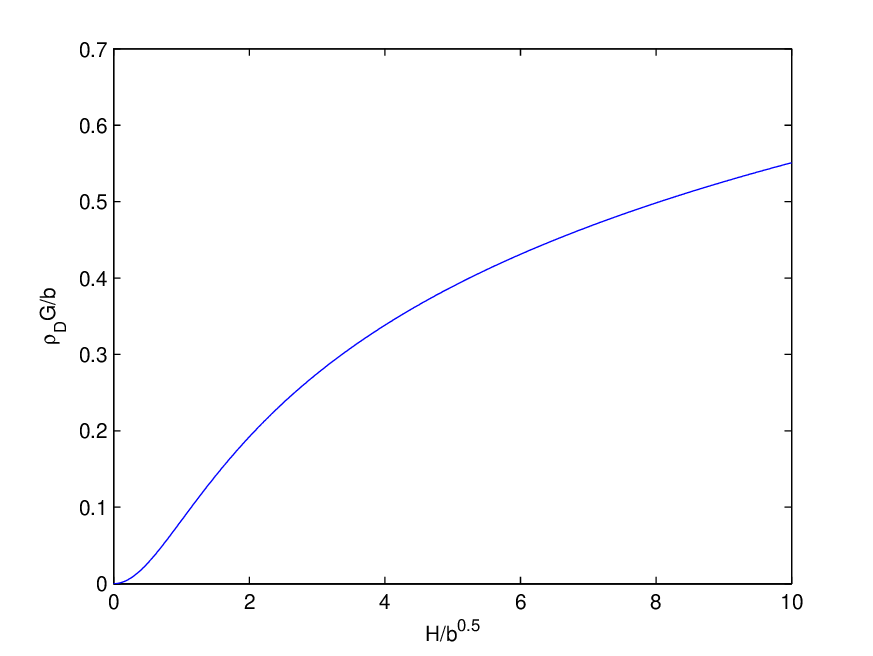}
\caption{\label{fig.1} The reduced dark energy density $\rho_{D}G/b$ vs. the parameter $H/\sqrt{b}$.}
\end{figure}
When $H/\sqrt{b}$ increases, the reduced dark energy density $\rho_DG/b$ also increases and $\lim_{H\rightarrow 0} \rho_{D}=0$.
As $H\rightarrow 0$ we have $R_h\rightarrow\infty$ which corresponds to the future era.

It is convenient to introduce the normalized density parameter of the matter $\Omega_m=\rho/(3M_P^2 H^2)$ and the normalized density parameter of dark energy $\Omega_D=\rho_D/(3M_P^2 H^2)$, where $M_P=1/\sqrt{8\pi G}$ is the reduced Planck mass. Then from Eq. (11) we obtain $\Omega_m+\Omega_D=1$. By the virtue of Eq. (12) one obtains the normalized density for the matter and the normalized density parameter of dark energy as follows:
\begin{equation}
\Omega_m=1-\frac{b}{H^2}\ln\left(\frac{H^2+b}{b}\right),~~~\Omega_D=\frac{b}{H^2}\ln\left(\frac{H^2+b}{b}\right).
\label{13}
\end{equation}
Making use of the dimensionless variable $x=H^2/b$, equations (13) become
\begin{equation}
\Omega_m=1-\frac{1}{x}\ln(1+x),~~~\Omega_D=\frac{1}{x}\ln(1+x).
\label{14}
\end{equation}
The normalized density parameter of the matter $\Omega_m$ and the normalized density parameter of dark energy $\Omega_D$ are depicted in Fig. 2.
\begin{figure}[h]
\includegraphics [height=2.0in,width=3.5in] {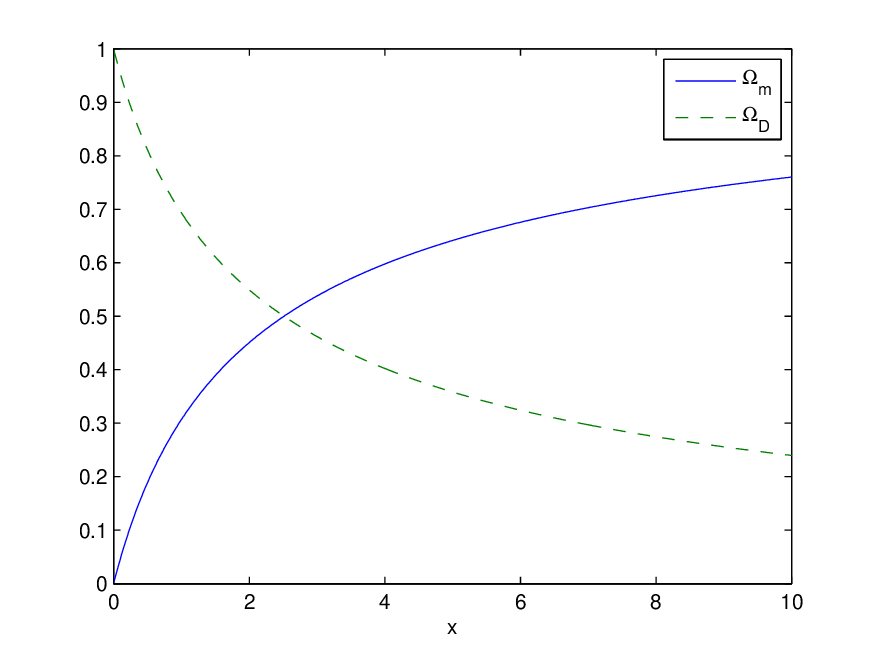}
\caption{\label{fig.2} The normalized density parameters $\Omega_m$ and $\Omega_D$ vs. $x=H^2/b$.}
\end{figure}
According to Fig. 2 as $x\rightarrow \infty$ ($H\rightarrow \infty$, $R_h\rightarrow 0$) we have $\Omega_m\rightarrow 1$ and $\Omega_D\rightarrow 0$ corresponding to the matter-dominated era. As $x\rightarrow 0$ ($H\rightarrow 0$, $R_h\rightarrow\infty$), $\Omega_D\rightarrow 1$ and $\Omega_m\rightarrow 0$ that corresponds to the dark-energy dominated epoch (the future era).
The Planck data show that $\Omega_{m0} \approx 0.315$ \cite{Aghanim} for the current era. The solution to Eq. (14) for $\Omega_{m0}=0.315$ is given by
\begin{equation}
x=\frac{1}{137}\left[-200 W_{-1}\left(-\frac{137}{200e^{137/200}}\right)-137\right]\approx 1.04282,
\label{15}
\end{equation}
where $W(z)$ is the Lambert function which obeys the equation $W\exp (W)=z$. The $W_{-1}(z)$ is the lower branch of $W(z)$ for $W(z)\leq -1$.
Then we obtain the entropy parameter
\begin{equation}
\alpha=\frac{bG}{\pi}=\frac{GH_0^2}{1.04282\pi}\approx 0.305~GH_0^2,
\label{16}
\end{equation}
where $H_0$ is the Hubble rate at the current time. To compute the EoS for dark energy $w_D$ we need the pressure $p_D$.
We assume that the dark energy and pressure obey the continuity equation (Eq. (5)) which gives
\begin{equation}
p_D=-\frac{\dot{\rho}_D}{3H}-\rho_D.
\label{17}
\end{equation}
By the virtue of Eqs. (12) and (17) one finds the equation for the pressure corresponding to dark energy
\begin{equation}
p_D=-\frac{b\dot{H}}{4\pi G(b+H^2)}-\frac{3b}{8\pi G}\ln\left(\frac{H^2+b}{b}\right).
\label{18}
\end{equation}
Making use of Eqs. (9), (10), (18) and the equation of state EoS for barotropic matter fluid $w=p/\rho$,  we obtain the pressure
\begin{equation}
p_D=\frac{3b(1+w)}{8\pi GH^2}\biggl [H^2-b\ln\left(\frac{H^2+b}{b}\right)\biggr ]-\frac{3b}{8\pi G}\ln\left(\frac{H^2+b}{b}\right).
\label{19}
\end{equation}
From Eqs. (12) and (19), we find the EoS for dark energy $w_D=p_D/\rho_D$,
\begin{equation}
w_D=\frac{1+w}{H^2}\left(\frac{H^2}{\ln\left(H^2/b+1\right)}-b\right)-1.
\label{20}
\end{equation}
By using the dimensionless variable $x=H^2/b$, Eq. (20) becomes
\begin{equation}
w_D=\frac{1+w}{x}\left(\frac{x}{\ln(x+1)}-1\right)-1.
\label{21}
\end{equation}
The EoS for dark energy $w_D$ versus $x$ is depicted in Fig. 3.
\begin{figure}[h]
\includegraphics [height=2.0in,width=3.7in] {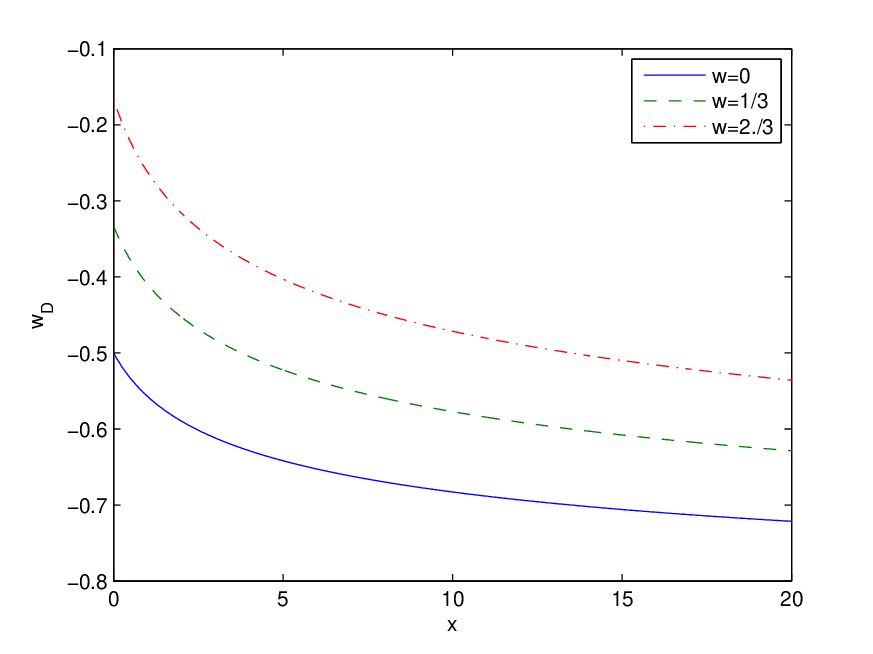}
\caption{\label{fig.3} The EoS for dark energy $w_D$ vs. the dimensionless variable $x=H^2/b$.}
\end{figure}
By the virtue of Eq. (21) one has $\lim_{x\rightarrow\infty}w_D=-1$ and $\lim_{x\rightarrow 0}w_D=(w-1)/2$. Thus, for the large Hubble parameter $H$ (the small $R_h$), the dark energy EoS is $w_D=-1$ which corresponds for the inflation era.

To analyse the observational data it is convenient to introduce the redshift $z = 1/a(t)-1$. Then using the continuity equation (5) and EoS $p=w\rho$, we obtain the density energy of the matter in the form
\begin{equation}
\rho=\rho_0(1+z)^{3(1+w)},
\label{22}
\end{equation}
where $\rho_0$ is the energy density of the matter at the present time. From Eqs. (10) and (22) we obtain the generalized Friedmann equation as follows:
\begin{equation}
H^2-b\ln\left(\frac{H^2+b}{b}\right)=\frac{8\pi G\rho_0}{3}(1+z)^{3(1+w)}.
\label{23}
\end{equation}
To plot the Hubble parameter versus redshift we find from Eq. (23) the redshift
\begin{equation}
z=\left[\frac{3}{8\pi\rho_0G}\left(H^2-b\ln\left(\frac{H^2+b}{b}\right)\right)\right]^{1/(3(1+w))}-1,
\label{24}
\end{equation}
and introducing dimensionless parameters $\bar{H}=H/\sqrt{G\rho_0}$, $\bar{b}=b/(G\rho_0)$, we represent Eq. (24) as
\begin{equation}
z=\left[\frac{3}{8\pi}\left(\bar{H}^2-\bar{b}\ln\left(\frac{\bar{H}^2+\bar{b}}{\bar{b}}\right)\right)\right]^{1/(3(1+w))}-1.
\label{25 }
\end{equation}
By the virtue of Eq. (25) we depicted the reduced Hubble parameter $\bar{H}$ versus redshift $z$ in Fig. 4.
\begin{figure}[h]
\includegraphics [height=2.0in,width=3.7in] {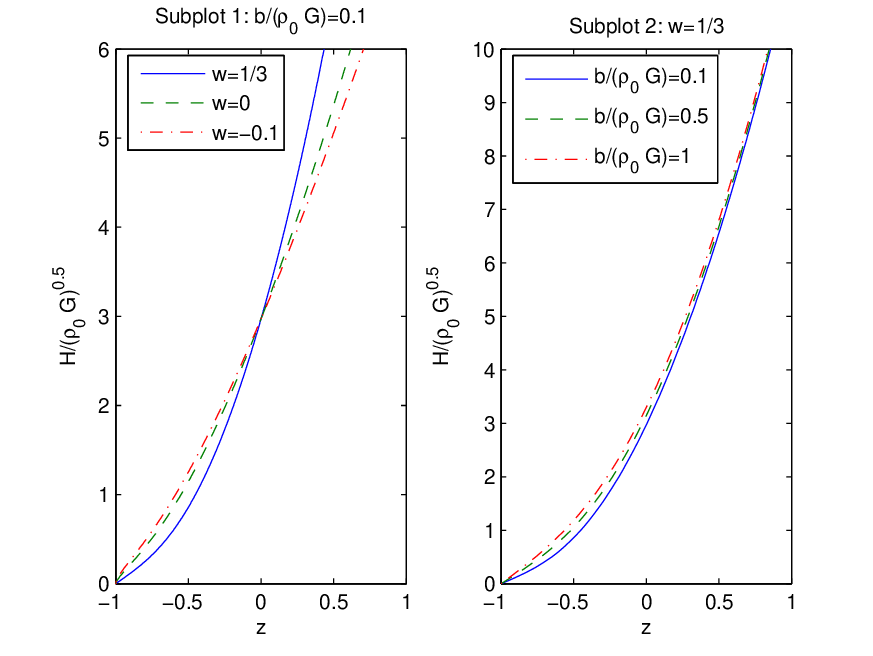}
\caption{\label{fig.4}The reduced Hubble rate $\bar{H}$ vs. redshift $z$.}
\end{figure}
In accordance with figure 4 when redshift $z$ increases the reduced Hubble parameter $\bar{H}$ also increases. According to the left panel of Fig. 4, when parameter $w$ increases at fixed $\bar{H}$, the redshift $z$ decreases at $z>0$. Right panel of Fig. 4 shows that when parameter $\bar{b}$ increases at fixed z, the reduced Hubble parameter $\bar{H}$ also increases. At $z=-1$ we have $H=0$.

Now we are going to fix the model parameter $w$ to agree with the Plank data. For this goal we consider
the deceleration parameter which is given by
\begin{equation}
q=-\frac{\ddot{a}a}{\dot{a}^2}=-1-\frac{\dot{H}}{H^2}.
\label{26}
\end{equation}
The acceleration phase of the universe occurs when $q<0$ and when $q>0$ the deceleration phase takes place. Making use of Eqs. (9), (22) and (26) we obtain the deceleration parameter as
\begin{equation}
q=\frac{4\pi G\rho_0(1+w)(H^2+b)}{H^4}\left(1+z\right)^{3(1+w)}-1.
\label{27}
\end{equation}
By the virtue of Eqs. (23) and (27) we find
\begin{equation}
q=\frac{3(1+w)(H^2+b)}{2H^4}\left(H^2-b\ln\left(\frac{H^2+b}{b}\right)\right)-1.
\label{28}
\end{equation}
Using dimensionless variable $x=H^2/b$, Eq. (28) becomes
\begin{equation}
q=\frac{3(1+w)(1+x)}{2x^2}\left(x-\ln(1+x)\right)-1.
\label{29}
\end{equation}
Making use of the value $x=H_0^2/b\approx 1.04282$ ($\alpha\approx H_0^2 G/(1.04282\pi)$), that gives the normalized density of the matter field at the current time $\Omega_{m0}\approx 0.315$, and the deceleration parameter $q_0\approx -0.535$ \cite{Aghanim}, we obtain the solution to Eq. (29) for the EoS parameter of the matter $w\approx -0.4976$.
We plotted the deceleration parameter $q$  versus $x$ in Fig. 5.
\begin{figure}[h]
\includegraphics [height=2.0in,width=3.7in] {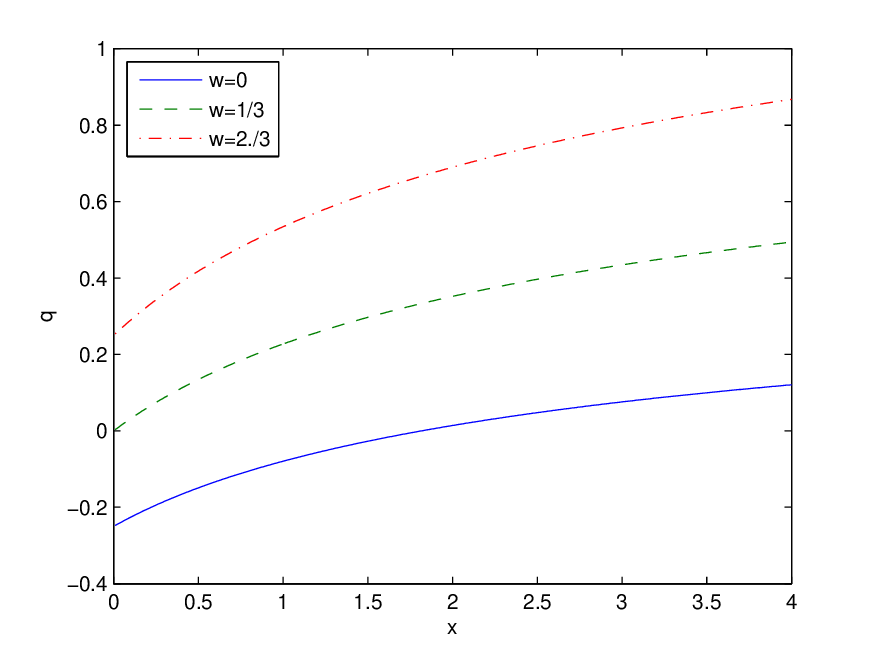}
\caption{\label{fig.5} The deceleration parameter $q$ vs. $x=H^2/b$ at $w=0, 1/3, 2/3$.}
\end{figure}
In accordance with Fig. 5 we have two phases, the universe acceleration and deceleration.
Taking into account Eq. (28), we obtain the asymptotic
\begin{equation}
\lim_{H\rightarrow \infty} q=\frac{3w+1}{2}.
\label{30}
\end{equation}
It follows from Eq. (30) that when $w>-1/3$ ($q>0$) at large $H$, the universe decelerates and the universe accelerates at $w<-1/3$. The calculated value $w\approx -0.4976$ is in agreement with this requirement.
Using Eq. (29) at $q=0$, we obtain the equation for the transition phase
\begin{equation}
w=\frac{2x^2}{3(1+x)\left(x-\ln(1+x)\right)}-1.
\label{31}
\end{equation}
We obtain limits of the EoS parameter $w$ at $x\rightarrow 0$ and $x\rightarrow \infty$ as follows:
\begin{equation}
\lim_{x\rightarrow 0}w=\frac{1}{3},~~~~\lim_{x\rightarrow \infty}w=-\frac{1}{3}.
\label{32}
\end{equation}
We plotted the EoS parameter for the matter $w$ versus $x$ in Fig. 6.
\begin{figure}[h]
\includegraphics [height=2.0in,width=3.7in] {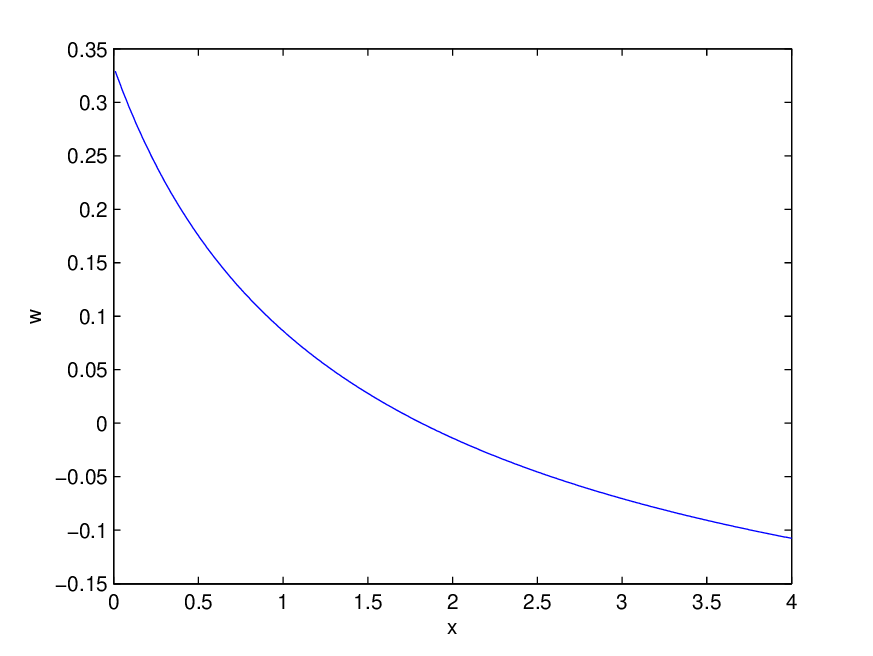}
\caption{\label{fig.6}  The EoS parameter for the matter $w$ vs. $x=H^2/b$ at $q=0$.}
\end{figure}
According to Fig. 6 when $x=H^2/b$ increases, the EoS parameter $w$ decreases. At large Hubble parameter $H$, for $q=0$, $w\rightarrow -1/3$ and at small $H$ we have $w\rightarrow 1/3$.

The entropy parameter is given by $\alpha=bG/\pi=GH^2/(x\pi)$. For the current era $x\approx 1$ and
$\alpha\approx GH_0^2/\pi$. Because $GH_0^2\ll 1$ we have for the current era $\alpha\ll 1$ and, therefore, quantum corrections to the Bekenstein--Hawking entropy $S_{BH}$, according to Eq. (8), are small.
When parameter $x$ is small, the entropy parameter $\alpha$ will be large and quantum effects are quite dominant. According to Fig. 2 this will happend when the normalized density parameter of dark energy  approaches to $\Omega_D\rightarrow 1$.

\section{F(T)-gravity from the R\'{e}nyi entropy}

In the theory of Teleparallel Equivalent to General Relativity (TEGR), the curvature is replaced by torsion. In this self-consistent theory of gravity dynamics is the same as in general relativity. In TEGR tetrad fields (vierbeins) define a basis that
describes the geometry of space-time. Vierbeins define a torsion tensor which is the source of gravity presenting the antisymmetric contribution of the Christoffel connection. The torsion scalar $T$ is constructed by the torsion tensor and defines the gravitational action. In $F(T)$ gravity the Lagrangian density of TEGR is modified by using an arbitrary function of the torsion scalar.
In the teleparallel theory of gravity, the Weitzenb\"{o}ck connection is used, and the field equations are the second-order.
The torsion field $T$ is given by \cite{Weitzen,Maluf}
\begin{equation}
T=S_\rho{}^{\mu\nu}T^\rho{}_{\mu\nu}.
\label{33}
\end{equation}
The superpotential $S_\rho{}^{\mu\nu}$ and the contortion tensor  $K^{\mu\nu}{}_\rho$  are
\[
S_\rho{}^{\mu\nu}=\frac{1}{2}\left(K^{\mu\nu}{}_\rho+\delta^\mu_\rho T^{\alpha\nu}{}_\alpha-\delta^\nu_\rho T^{\alpha\mu}{}_\alpha\right),
\]
\begin{equation}
K^{\mu\nu}{}_\rho=-\frac{1}{2}\left(T^{\mu\nu}{}_\rho-T^{\nu\mu}{}_\rho-T_\rho{}^{\mu\nu}\right),
\label{34}
\end{equation}
and the torsion tensor is defined as
\begin{equation}
T^\rho{}_{\mu\nu}=e^\rho_{i}\left(\partial_\mu e^i_{\nu}-\partial_\nu e^i_{\mu}\right),
\label{35}
\end{equation}
where $e^i_{\nu}$ ($i = 0, 1, 2, 3$) is  a vierbein field. In the flat metric of the tangent spacetime $\eta_{ij}$, the metric tensor is $g_{\mu\nu}=\eta_{ij}e^i_\mu e^j_\nu$. In FLRW metric (1), the vierbein field is $e^i_\mu=\mbox{diag}(1,a,a,a)$ and the torsion scalar becomes $T=-6H^2$. The variation of the action with respect to $e^i_\mu$ with the Lagrangian $F(T)$ gives the equation \cite{Bengochea}
\begin{equation}
\frac{1}{6}\left[F(T)-2TF'(T)\right]=\left(\frac{8\pi G}{3}\right)\rho.
\label{36}
\end{equation}
Making use of Eqs. (10) and (36) and $T=-6H^2$ we obtain
\begin{equation}
F'(T)-\frac{F(T)}{2T}=\frac{1}{2}+\frac{3b}{T}\ln\left(1-\frac{T}{6b}\right).
\label{37}
\end{equation}
By integrating Eq. (37) we find the equation as follows:
\begin{equation}
F(T)=T-6b\ln\left(1-\frac{T}{6b}\right)+2\sqrt{-6bT}\arctan\left(\sqrt{-\frac{T}{6b}}\right),
\label{38}
\end{equation}
where we use the integration constant to be $C=0$. We have used the relation $i~\mbox{tanh}^{-1}(ix)=-\arctan(x)$ because $T=-6H^2<0$. Some teleparallel gravity models were studied in \cite{Wu,Saridakis}. Thus, we showed that the entropic cosmology with the R\'{e}nyi entropy (7) is equivalent to a cosmology based on the teleparallel gravity with the function (38).

\section{Conclusion}

We have studied the entropic cosmology with the R\'{e}nyi entropy $S_R=(1/\alpha)\ln(1+\alpha S_{BH})$ that describes the dark energy and leads to the current acceleration of the universe. The spatial flat FLRW universe and the matter barotropic perfect fluid are implied.
By the virtue of the first law of apparent horizon thermodynamics we have obtained the modified Friedmann equations which include the density of dark energy. We assumed that the dark energy density $\rho_D$ and pressure  $p_D$ obey the continuity equation (the conservation law).
The EoS $w_D=p_D/\rho_D$ has been computed with $\lim_{H\rightarrow\infty}w_D=-1$ which shows that at $R_h\rightarrow 0$ the de Sitter space-time occurs and the inflation of the universe takes place.  In the model under consideration the universe may have two phases, acceleration and deceleration, due to the dark energy. We have showed that at the entropy parameter $\alpha\approx 0.305 GH_0^2$ and $w=-0.4976$ the deceleration parameter has the value $q_0\approx -0.535$ and the normalized density parameter of the matter is $\Omega_{m0}\approx 0.315$ that are in agrement with the Planck data at the current epoch \cite{Aghanim}. It was shown that entropic cosmology studied can be considered as the cosmology
based on the teleparallel gravity with the function $F(T)$ obtained. The Hubble parameters are in approximate agreement with the observational Hubble data for $0.07\leq z \leq 1.75$ at the entropy parameter $\alpha=bG/\pi\approx 0.305~GH_0^2$ (see Appendix).

\section{Appendix: The Hubble parameter and observational data}

Making use of Eq. (10) and the energy density
\begin{equation}
\rho=\rho_m+\rho_r=\rho_{m0}(1+z)^3+\rho_{r0}(1+z)^4,
\label{39}
\end{equation}
where $\rho_m$ is the energy density of the non-relativistic matter in the form of dust ($w=0$) and $\rho_r$ is the radiation energy density ($w=1/3$), we obtain
\begin{equation}
H^2(z)-b\ln\left(\frac{H^2(z)}{b}+1\right)=H_0^2\Omega_{m0}(1+z)^3+H_0^2\Omega_{r0}(1+z)^4.
\label{40}
\end{equation}
One can neglect the contribution of radiation energy density for small redshifts because $\Omega_{r0}\approx 10^{-4}$.
In accordance with Eq. (40) at $z=-1$ we have the value $H(-1)=0$. By the virtue of Eq. (40) with the current Hubble parameter $H_0=67~\mbox{km/Mpc/s}$ and $\Omega_{m0}=0.315$, we depicted the Hubble parameter $H(z)$ in units of km/Mpc/s in Fig. 7.
\begin{figure}[h]
\includegraphics [height=2.0in,width=3.7in] {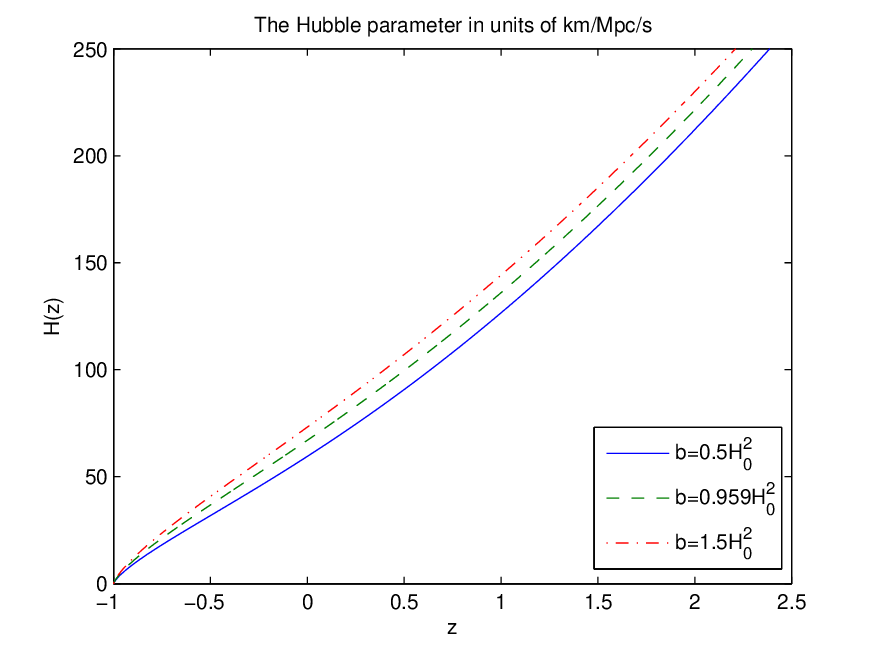}
\caption{\label{fig.7} The Hubble parameter $H(z)$ in units of km/Mpc/sec vs. $z$.}
\end{figure}
 According to Fig. 7, the values of Hubble parameters are in approximate agreement with the observational Hubble data for $0.07\leq z \leq 1.75$
 \cite{Zhang,Moresco,Simon,Stern,Loubser,Moresco1,Jimenez,Jiao,Tomasetti,Moresco2, Ratsimbazafy} at $b=\alpha\pi/G\approx0.959~H_0^2$ ($\alpha\approx0.305~GH_0^2$) which gives the correct value for the normalized density parameter of the matter at the current era. Some observational Hubble data for $0.07\leq z \leq 1.75$ are represented in Table 1.
 \begin{table}[ht]
\caption{Observational Hubble data for $0.07\leq z \leq 1.75$. The Hubble parameter $H_{OHD}$ is in units of km/Mpc/s.}
\centering
\begin{tabular}{c c c c c c c c c c  c c c c}\\[1ex]
\hline
$z$ & 0.07 & 0.18 & 0.24  & 0.429 & 0.45 & 0.48 & 0.593 & 0.875 & 1.3 & 1.75  \\[0.5ex]
\hline
$H_{OHD}$ & 69 & 75 & 79.7 & 91.8 & 92.8 & 97 & 104 & 125 & 168 & 202   \\[0.5ex]
\hline
Ref. & [50] & [55] & [52] & [51] & [51] & [53] & [55] & [55]  & [52] & [52]   \\[0.5ex]
\hline
\end{tabular}
\end{table}

Introducing the dimensionless parameter $E(z)=H(z)/H_0$ and making use of parameter $b\approx0.959H_0^2$ which gives the normalized density parameter $\Omega_{m0}=0.315$, we represent Eq. (40) as follows:
\begin{equation}
E(z)^2(z)-0.959\ln\left(\frac{E^2(z)}{0.959}+1\right)=0.315(1+z)^3+10^{-4}(1+z)^4.
\label{41}
\end{equation}
Solving Eq. (41) and by the virtue of relation $H(z)=E(z)H_0$ with $H_0=67$ km/Mpc/s we obtain the Hubble parameters for the interval $0.07\leq z \leq 1.7$  represented in Table 2. We include also in Table 2 the relative percentage deviation $R=100(H-H_{OHD})/H_{OHD}$ which shows the deviation of observational Hubble parameters from the predicted Hubble parameters.
 \begin{table}[ht]
\caption{The Hubble parameters (in km/Mpc/s) for $0.07\leq z \leq 1.75$, calculated from Eq. (41) with relation $H(z)=E(z)H_0$, and the relative percentage deviation $R$.}
\centering
\begin{tabular}{c c c c c c c c c c  c c c c}\\[1ex]
\hline
$z$ & 0.07 & 0.18 & 0.24  & 0.429 & 0.45 & 0.48 &0.593 & 0.875 & 1.3  & 1.75  \\[0.5ex]
\hline
$H$ & 71.4 & 78.4 & 82.2 & 94.7 & 95 & 98.2 & 106.0 & 125 & 159.8 & 198.5   \\[0.5ex]

\hline
$R$ & 3.43 & 4.47 & 3.17 & 3.20 & 2.37 & 1.25 & 1.96 & 0 & -4.88 & -1.76   \\[0.5ex]
\hline
\end{tabular}
\end{table}

Table 2 shows that the deviation of predicted values of Hubble parameters from the observational Hubble data are small and the relative percentage deviations $R=100(H-H_{OHD})/H_{OHD}$ are within 5 percents.


\end{document}